\documentclass{article}
\usepackage{PRIMEarxiv}

\usepackage[utf8]{inputenc} % allow utf-8 input
\usepackage[T1]{fontenc}    % use 8-bit T1 fonts
\usepackage{hyperref}       % hyperlinks
\usepackage{url}            % simple URL typesetting
\usepackage{booktabs}       % professional-quality tables
\usepackage{amsfonts}       % blackboard math symbols
\usepackage{nicefrac}       % compact symbols for 1/2, etc.
\usepackage{microtype}      % microtypography
\usepackage{lipsum}
\usepackage{graphicx}
\graphicspath{{media/}}     % organize your images and other figures under media/ folder

\usepackage{amsmath}
\usepackage{amssymb}
\usepackage{booktabs}
\usepackage{multirow}
\usepackage{enumitem}
\usepackage{caption}
\usepackage{bbold}
\usepackage{ragged2e}
\usepackage{subcaption}

\usepackage[capitalize]{cleveref}
\crefname{section}{Sec.}{Secs.}
\Crefname{section}{Section}{Sections}
\Crefname{table}{Table}{Tables}
\crefname{table}{Tab.}{Tabs.}

\usepackage{overpic}
\usepackage{enumitem} %< control spacing in itemize/enumerate/...
\usepackage{overpic} %< add raw math symbols to figures
\usepackage{color}
\usepackage{microtype} %< hardcore text layout optimization (ONLY UPDATE ~DEADLINE)
\definecolor{turquoise}{cmyk}{0.65,0,0.1,0.3}
\definecolor{purple}{rgb}{0.65,0,0.65}
\definecolor{dark_green}{rgb}{0, 0.5, 0}
\definecolor{orange}{rgb}{0.8, 0.6, 0.2}
\definecolor{red}{rgb}{0.8, 0.2, 0.2}
\definecolor{darkred}{rgb}{0.6, 0.1, 0.05}
\definecolor{blueish}{rgb}{0.0, 0.3, .6}
\definecolor{light_gray}{rgb}{0.7, 0.7, .7}
\definecolor{pink}{rgb}{1, 0, 1}
\definecolor{greyblue}{rgb}{0.25, 0.25, 1}

\newcommand{\Fig}[1]{Fig.~\ref{fig:#1}}

\newcommand{\Tab}[1]{Tab.~\ref{tab:#1}}

\newcommand{\Sec}[1]{Sec.~\ref{sec:#1}}

\usepackage{blindtext}

\renewcommand{\paragraph}[1]{\vspace{1em}\noindent\textbf{#1}.}
\usepackage{bbold}
\renewcommand\vec{\mathbf}
\newcommand{\mat}{\mathbf}
\usepackage{wrapfig}
\usepackage{multirow}

\title{Recurrent Variational Network: A Deep Learning Inverse Problem Solver
applied to the task of Accelerated MRI Reconstruction
}

\author{
 George Yiasemis \\
  Netherlands Cancer Institute,\\
  Amsterdam, 1066 CX,\\
  the Netherlands \\
  \texttt{g.yiasemis@nki.nl} \\
   \And
 Jan-Jakob Sonke \\
  Netherlands Cancer Institute,\\
  Amsterdam, 1066 CX,\\
  the Netherlands \\
  \texttt{j.sonke@nki.nl} \\
   \And
  Clarisa Sánchez \\
  qurAI group\\
  University of Amsterdam,\\
  Amsterdam, 1012 WX, \\
  the Netherlands \\
  \texttt{c.i.sanchezgutierrez@uva.nl} \\
  \And
   Jonas Teuwen \\
   Netherlands Cancer Institute,\\
  Amsterdam, 1066 CX,\\
  the Netherlands \\
  \texttt{j.teuwen@nki.nl} \\
}

\begin{document}
\maketitle

\begin{abstract}

Magnetic Resonance Imaging can produce detailed images of the anatomy and physiology of the human body that can assist doctors in diagnosing and treating pathologies such as tumours. However, MRI suffers from very long acquisition times that make it susceptible to patient motion artifacts and limit its potential to deliver dynamic treatments. Conventional approaches such as Parallel Imaging and Compressed Sensing allow for an increase in MRI acquisition speed by reconstructing MR images from sub-sampled MRI data acquired using multiple receiver coils.  Recent advancements in Deep Learning combined with Parallel Imaging and Compressed Sensing techniques have the potential to produce high-fidelity reconstructions from highly accelerated MRI data. In this work we present a novel Deep Learning-based Inverse Problem solver applied to the task of Accelerated MRI Reconstruction, called the Recurrent Variational Network (RecurrentVarNet), by exploiting the properties of Convolutional Recurrent Neural Networks and unrolled algorithms for solving Inverse Problems. The RecurrentVarNet consists of multiple recurrent blocks, each responsible for one iteration of the unrolled variational optimization scheme for solving the inverse problem of multi-coil Accelerated MRI Reconstruction. Contrary to traditional approaches, the optimization steps are performed in the observation domain ($k$-space) instead of the image domain. Each block of the RecurrentVarNet refines the observed $k$-space and comprises a data consistency term and a recurrent unit which takes as input a learned hidden state and the prediction of the previous block. Our proposed method achieves new state of the art qualitative and quantitative reconstruction results on 5-fold and 10-fold accelerated data from a public multi-coil brain dataset, outperforming previous conventional and deep learning-based approaches. Our code is publicly available at \href{https://github.com/NKI-AI/direct}{\color{magenta}{https://github.com/NKI-AI/direct}}.
\end{abstract}

% keywords can be removed
\section{Introduction}
\label{sec:intro}
\begin{figure}[t]
\begin{center}
\includegraphics[width=\linewidth]{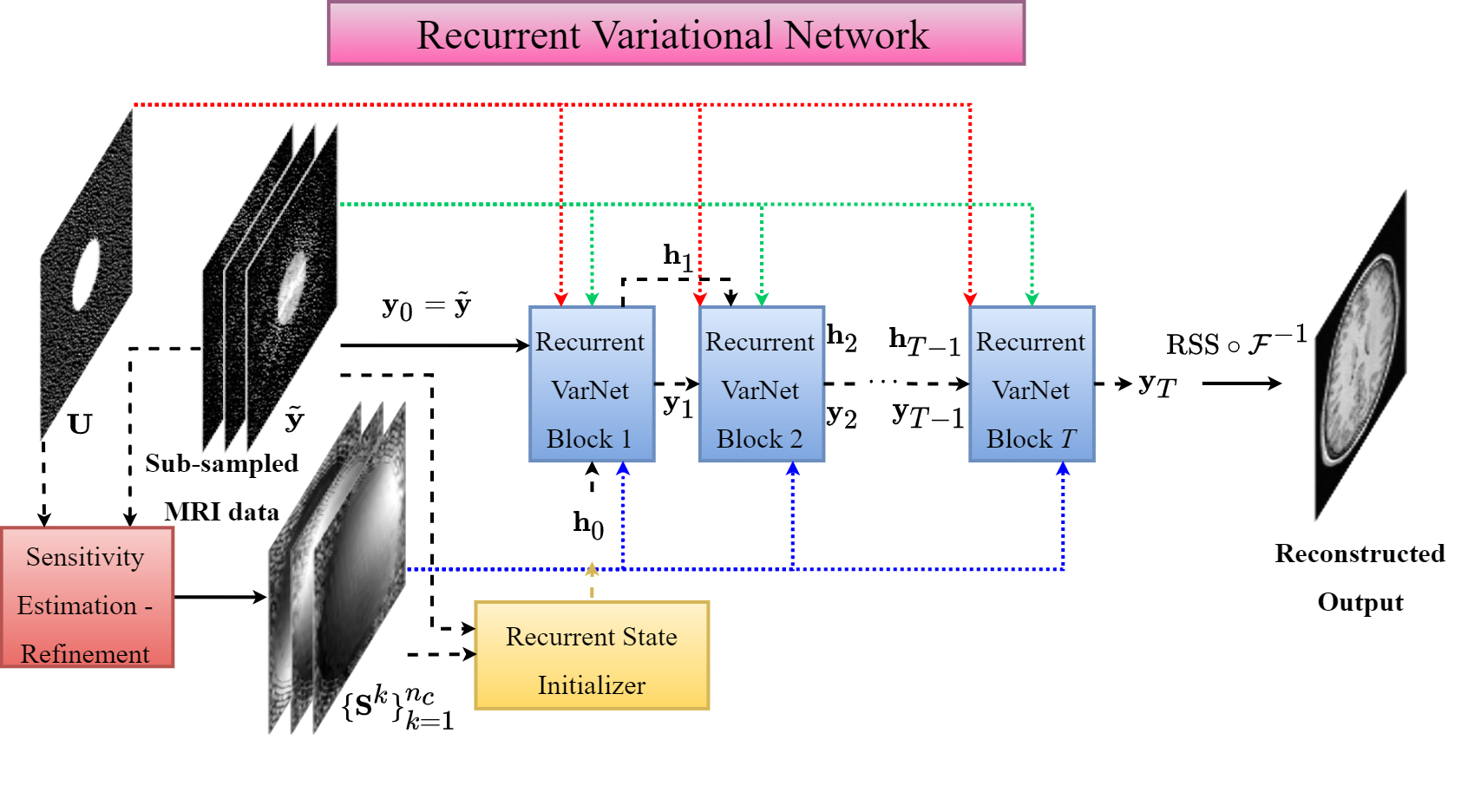}
\end{center}
\caption{Overview of our proposed framework, Recurrent Variational Network (\textbf{RecurrentVarNet}), applied to the task of multi-coil Accelerated MRI Reconstruction. Our model takes as input sub-sampled MRI data from multiple coils which are refined following an iterative gradient descent-like optimization scheme of $T$ time-steps in the $k$-space, and outputs an estimate of the ground truth reconstructed image.}
\label{fig:teaser}
\end{figure}

Magnetic Resonance Imaging (MRI) is one of the most vastly used imaging modalities in medicine. Since MRI allows for production of highly detailed anatomical images of the human body, it is often employed for disease detection, prognosis, and treatment monitoring and guidance. The facts that MRI is non-invasive and that it does not involve exposure to ionizing radiation have played a pivotal role in making it a popular imaging technique. 

However, MRI is still bounded by very lengthy acquisition times induced by the fact that the MRI scanner reads data in the frequency domain ($k$-space) sequentially in time. This usually causes patient discomfort and motion artifacts, and it also limits its use for delivery of dynamic treatments such as image-guided radiotherapy. Therefore, reducing the scanning time in MRI, often referred to as \textit{accelerating MRI} \cite{zbontar2019fastmri}, not only could assist in reducing medical costs and patients' distress, but could also make dynamic treatments feasible. Conventional approaches for accelerating MRI include Parallel Imaging (PI) \cite{Larkman2007ParallelMR, encodingandreconpi, lustigCS2}, Compressed Sensing (CS) \cite{lustigCS, 1614066}, and the combination of these two methods (PI-CS) \cite{6235721}.

In contrast to traditional MRI acquisitions, in PI multiple, instead of one, radio-frequency receiver coils are employed to simultaneously obtain reduced sets of measurements to shorten the acquisition time. Coil sensitivity maps are required to be known since each coil is sensitive to a different part of the volume according to its position. Classical approaches for estimating these coil sensitivity maps include E-SPIRiT \cite{Uecker2014ESPIRiTanEA}, SENSE \cite{isense, senseencoding}, GRAPPA \cite{grappa}, and SMASH \cite{smash}.

Besides PI, CS techniques aim in accelerating the MRI acquisition by sub-sampling the $k$-space, that is, acquiring sparse or partial raw $k$-space measurements. Unfortunately, this is accompanied by a degradation of image quality, known as aliasing artifacts, which include low-resolution images, blurring, or in-folding artifacts \cite{doi:10.1148/rg.313105115}. This is attributed to the fact that this violates the Nyquist-Shannon sampling criterion \cite{grenander1959probability, 1697831}. In CS, an acceleration factor $R$ is chosen which determines the magnitude of the acceleration, and is referred to as $R$-fold acceleration.  $R$ is calculated as the ratio of the number of $k$-space measurements needed for a full scan over the number of the acquired measurements.

Recent advancements in Deep Learning (DL) and specifically in Convolutional Neural Networks (CNNs), have led to their application in many fields including solving inverse problems arising in imaging, such as Accelerated MRI Reconstruction tasks. Combined with PI-CS, DL-based methods can outperform conventional PI and CS approaches by producing reconstructed MR images with less artifacts \cite{cstoAI}. Additionally, in 2018-19 the fastMRI \cite{zbontar2019fastmri, fastmri2021} and Calgary-Campinas \cite{beauferris2020multichannel} challenges, have provided the MRI community with publicly available raw MRI data enabling the development of multiple DL state of the art baselines for producing faithful reconstructed images from sub-sampled $k$-space measurements \cite{kikinet, ramzi2021xpdnet, sriram2020endtoend, LONNING201964, Jun_2021_CVPR}. 

Motivated by the need to accelerate the MRI acquisition even further, in this work we propose a novel CNN and recurrent-based DL inverse problem solver inspired by the Variational Network \cite{hammernik2017learning} and the Recurrent Inference Machine \cite{putzky2017recurrent} approaches which we apply on the task of reconstructing images from accelerated MRI data. We call our proposed method the \textbf{Recurrent Variational Network}.
 
\paragraph{Contributions}
\begin{itemize}
    \item We propose the Recurrent Variational Network, a novel Deep Learning architecture for solving imaging inverse problems. We applied our method to the multi-coil accelerated MRI reconstruction task.
    \item We are the first to present a recurrent-based DL imaging inverse problem solver that performs iterative optimization in the measurements ($k$-space) domain.
    % , instead of the image domain.
    \item Our quantitative and qualitative experiments demonstrate that our model achieved state of the art results on the Calgary-Campinas public brain dataset outperforming current baselines.
\end{itemize}

\section{Background and Related work}
\label{sec:related}

\subsection{MRI Acquisition}
\label{sec:subsec2.1}
In clinical settings, the acquisition of MRI measurements by the MR scanner, also known as $k$-space, is performed in the frequency or Fourier domain. Let $n \, = \, n_{x}\times n_{y}$ denote the spatial size of the data. In the case of single-coil acquisition, the equation between the underlying spatial ground truth image $\vec{x} \, \in \,  \mathbb{C}^{n}$ and the $k$-space $\vec{y} \, \in \, \mathbb{C}^{n}$ is given by
\begin{equation}
    \vec{y} \, = \, \mathcal{F}(\vec{x}) \, + \, \vec{e},
\end{equation}
where $\mathcal{F}$ denotes the two-dimensional Fourier transform and $\vec{e} \, \in \, \mathbb{C}^{n}$ denotes some  measurement noise, usually assumed additive and normally distributed.

\subsection{Accelerated MRI Acquisition}
\label{sec:subsec2.2}

In Parallel Imaging, multiple ($n_c$) receiver coils are placed around the subject to speed up the acquisition. The acquired measurements $\vec{y} \, = \, \{\vec{y}^k\}_{k=1}^{n_c}$ in the case of multi-coil acquisition can be described as
\begin{equation}
    \vec{y}^k \, = \, \mathcal{F}(\mat{S}^k\vec{x}) \, + \, \vec{e}^k, \quad k\,=\,1,\,2,\,...,\,n_{c},
    \label{eq:multicoil}
\end{equation}
where $\vec{e}^k$ denotes measurement noise from the $k$-th coil and $\mat{S}^k\,\in\, \mathbb{C}^{n\times n}$ is called the sensitivity map of the $k$-th coil. The sensitivity maps encode the spatial sensitivity of their respective coil and they can be expressed with diagonal matrices. They are usually normalized such that
\begin{equation}
    \sum_{k=1}^{n_c} {\mat{S}^k}^{*}{\mat{S}^k}\,=\, \mat{I}_n.
\end{equation}
To accelerate the MRI acquisition, in CS settings the $k$-space is sub-sampled. The sub-sampled $k$-space measurements can be defined in terms of the fully-sampled $k$-space measurements as follows:
\begin{equation}
    \tilde{\vec{y}}^k \, = \, \mat{U}\vec{y}^k \, = \, \mat{U}\mathcal{F}(\mat{S}^k\vec{x}) \, + \,\tilde{\vec{e}}^k,\quad k\,=\,1,\,2,\,...,\,n_{c}
    \label{eq:multicoil_subsampled}
\end{equation}
where $\mat{U}\,\in\,\{0,\,1\}^{n}$ is a sub-sampling mask operator and is dependent on the acceleration factor $R$. Note that the same sub-sampling mask is applied to each coil data.

Sensitivity maps can be estimated by fully-sampling the center region of the $k$-space. This is known in the literature as the auto-calibration signal (ACS) which includes the low-frequency samples. We denote as $\mat{U}_{\text{ACS}}$ the ACS operator which takes as input $k$-space data and outputs the auto-calibration signal. An estimate for the sensitivity map of the $k$-th coil can be obtained as
\begin{equation}
    \tilde{\mat{S}}^k\, = \,\mathcal{F}^{-1} (\mat{U}_\text{ACS} \tilde{\vec{y}}^k ) \oslash \text{RSS} \Big( \big \{\mathcal{F}^{-1} ( \mat{U}_\text{ACS}\tilde{\vec{y}}^l )\big\}_{l=1}^{n_c}\Big),
    \label{eq:sensemaps}
\end{equation}
where $\oslash$ denotes the element-wise (Hadamard) division and $\text{RSS}:\mathbb{C}^{n\times n_c}\rightarrow\mathbb{R}^{n}$ the root-sum-of-squares operator \cite{LARSSON2003121} defined as
\begin{equation}
    \text{RSS}(\vec{x}^1\,...,\,\vec{x}^{n_c})\,=\,\bigg(\sum_{k=1}^{n_c}|\vec{x}^k|^{2}\bigg)^{1/2}.
    \label{eq:rss}
\end{equation}
Note that $\mat{U}(\vec{y})\,=\,\mat{U}_{\text{ACS}}(\vec{y})$ for the samples of the ACS region.
 
\subsection{Accelerated MRI Reconstruction as an Inverse Problem}
\label{sec:subsec2.3}

Reconstructing MR images from sub-sampled data is an inverse problem with \eqref{eq:multicoil_subsampled} posing as the forward model. Since \eqref{eq:multicoil_subsampled} is ill-posed \cite{ill-posed}, an explicit solution for the image $\vec{x}$ is hard to obtain. In general, \eqref{eq:multicoil_subsampled} can be treated as a variational optimization problem \cite{hammernik2017learning, Kaipio:1338003}, that is, solving
\begin{equation}
        \hat{\vec{x}} \, = \, \arg \min_{\vec{z} \, \in \, \mathbb{C}^{n} } \, \sum_{k=1}^{n_{c}} \mathcal{L} \big( \mat{U}\mathcal{F}(\mat{S}^k\vec{z}), \, \tilde{\vec{y}}^{k} \big) \, + \, \lambda \, \mathcal{C}(\vec{z})
         \label{eq:varproblem}
\end{equation}
where $\mathcal{L} : \mathbb{C}^{n\times n_c} \times \mathbb{C}^{n\times n_c} \rightarrow \mathbb{R}^{+}$ and $\mathcal{C} : \mathbb{C}^{n} \rightarrow \mathbb{R}$ denote convex data fidelity and regularization functionals, respectively, and $\lambda \, > \, 0$ the regularization parameter. 

By defining an expand operator $\mathcal{E} : \mathbb{C}^{n} \times \mathbb{C}^{n\times n_c}  \rightarrow  \mathbb{C}^{n\times n_c}$ that takes as input an image and outputs the individual coil images and a reduce operator $\mathcal{R} : \mathbb{C}^{n\times n_c} \times \mathbb{C}^{n\times n_c}  \rightarrow  \mathbb{C}^{n}$ that combines the individual coil images into one image as
\begin{equation}
    \begin{gathered}
        \mathcal{E}(\vec{z}) \, = \, (\mat{S}^1\vec{z},\,...,\,  \mat{S}^{n_c}\vec{z})\,=\,(\vec{z}^1,\,...,\,\vec{z}^{n_c}) \\
        \mathcal{R}(\vec{z}^1,\,...,\,\vec{z}^{n_c}) \, = \, \sum_{k=1}^{n_c}{\mat{S}^k}^*\vec{z}^k,
    \end{gathered}
\end{equation}
we can define as the linear forward operator of multi-coil accelerated MRI reconstruction $\mathcal{A} : \mathbb{C}^{n} \rightarrow \mathbb{C}^{n\times n_c}$ and its adjoint (backward) operator $\mathcal{A}^{*} : \mathbb{C}^{n\times n_c} \rightarrow \mathbb{C}^{n}$ as composition of operators:
\begin{equation}
            \mathcal{A} \, = \, \mat{U} \, \circ \, \mathcal{F} \, \circ \, \mathcal{E}, \quad
            \mathcal{A}^* \, = \, \mathcal{R}  \, \circ \, \mathcal{F}^{-1} \, \circ \, \mat{U}, 
\end{equation}
where $\mathcal{F}$, $\mathcal{F}^{-1}$ and $\mat{U}$ are applied element-wise. 

Let $\tilde{\vec{y}}\,=\,(\tilde{\vec{y}}^1,\,...,\,\tilde{\vec{y}}^{n_c})$, then, we can define \eqref{eq:varproblem} in a more compact notation:
\begin{equation}
     \hat{\vec{x}}\, = \,  \arg \min_{\vec{z} \, \in \, \mathbb{C}^{n} } \, \mathcal{L}\big(\mathcal{A}(\vec{z}), \, \tilde{\vec{y}}\big) \, + \, \lambda \, \mathcal{C}(\vec{z}).
     \label{eq:varproblemcompact}
\end{equation}
Note that the operators $\mathcal{E}$ and $\mathcal{R}$ take also as inputs the coil sensitivity maps which we omit to save space. 

In this work we solve a regularised least-squares variational problem, that is, replacing $\mathcal{L}$ in \eqref{eq:varproblemcompact} with the squared $\mathcal{L}_2$-norm:
\begin{equation}
     \hat{\vec{x}}\, = \,  \arg \min_{\vec{z} \, \in \, \mathbb{C}^{n} } \,\frac{1}{2} \, ||\mathcal{A}(\vec{z}) \, - \, \tilde{\vec{y}}||_2^2 \, + \, \lambda \, \mathcal{C}(\vec{z}).
     \label{eq:least}
\end{equation}
An approximate solution to \eqref{eq:least} can be acquired by unrolling a gradient descent scheme \cite{hammernik2017learning} solved over $T$ iterations
\begin{equation}
    \vec{z}_{t+1} \, = \, \vec{z}_{t} \, - \, \alpha_{t+1} \,\bigg( \mathcal{A}^{*}\big(\mathcal{A}(\vec{z}_{t}) \, - \, \tilde{\vec{y}}\big) \, + \, \lambda \, \vec{\nabla} \mathcal{C}(\vec{z}_{t})\bigg)
    \label{eq:gd_unrolled}
\end{equation}
where $t \, = \, 0,\,1,\, ... ,\,T\,-\,1$, and $\alpha_{t+1}\,>\,0$ is the step-size at iteration $t$ and $\vec{z}_{0}$ is a suitable initial guess.

\subsection{Deep Accelerated MRI Reconstruction}
\label{sec:subsec2.4}

With the advent of the involvement of Deep Learning in accelerated MRI reconstruction tasks, a plethora of algorithms have been deployed with some reaching state of the art performance \cite{fastmri2021, beauferris2020multichannel}. These methods can be divided into three sub-groups:
\begin{itemize}[leftmargin=*]
\setlength\itemsep{-.3em}
    \item Image domain-learning models which operate exclusively in the image domain \cite{hammernik2017learning, zbontar2019fastmri,Jun_2021_CVPR, LONNING201964}.
    \item $k$-space domain-learning models which operate exclusively in the $k$-space domain \cite{zhangmultichannel, scanspecifickspace, kim2019loraki}.
    \item Hybrid (image and $k$-space) domain-learning models which alternate between image and $k$-space domains \cite{2018lpdnet, souza2019dualdomain, kikinet,ramzi2021xpdnet, sriram2020endtoend}.
\end{itemize}

In this paper we base our work on two approaches: the End-to-End Variational Network (E2EVarNet) \cite{sriram2020endtoend} and the Recurrent Inference Machine (RIM) \cite{putzky2017recurrent, LONNING201964}. The former is a hybrid domain-learning model which builds upon Variational Networks \cite{hammernik2017learning} by using a CNN-based neural network in the image domain to perform an iterative optimization scheme in the $k$-space domain by adapting \eqref{eq:least}. The latter, is an image domain-learning model that solves \eqref{eq:least} by treating it as a maximum-a-posteriori (MAP) estimation problem \cite{Kaipio:1338003}. By employing a sequence of alternating CNNs and Convolutional Gated Recurrent Units (ConvGRUs), RIM learns to predict an incremental update after each iteration using as inputs the previous image prediction, a hidden state, and the gradient of the negative log-likelihood distribution.

In \Sec{method} we extensively describe the architecture of our proposed method which is a hybrid domain-learning model, as it performs optimization in the measurements ($k$-space) domain using a recurrent CNN-based network which, however, operates in the image domain.
\section{Method}
\label{sec:method}

In this work we introduce the Recurrent Variational Network (RecurrentVarNet), which is an end-to-end inverse problem solver framework that performs a gradient descent-like optimization in the measurements domain using convolutional recurrent neural networks. In this paper we particularized RecurrentVarNet to the task of Accelerated MRI Reconstruction following the next steps: 

\noindent
First, we replaced the need of explicitly defining a regularization term $\mathcal{C}$ in \eqref{eq:least} and computing its gradient in \eqref{eq:gd_unrolled} with a CNN-based image-refinement model $\mathcal{G}_{\theta_{t+1}}$ at each layer $t\,=\,0,\,1,\,...,\,T\,-\,1$:
\begin{equation}
    \vec{z}_{t+1} \, = \, \vec{z}_{t} \, - \, \alpha_{t+1} \, \mathcal{A}^{*}\big(\mathcal{A}(\vec{z}_{t}) \, - \, \tilde{\vec{y}}\big) \, + \, \mathcal{G}_{\theta_{t+1}} (\vec{z}_{t}).
    \label{eq:varnet}
\end{equation}
\noindent
Next, to perform optimization in the $k$-space, we applied the operator $\mathcal{F} \, \circ \, \mathcal{E}$ on both sides of \eqref{eq:varnet}:
\begin{equation}
    \begin{split}
        \vec{y}_{t+1} \, = \, \vec{y}_{t} \,- \, \alpha_{t+1} \,\mat{U}\big(\vec{y}_{t}\,-\,\tilde{\vec{y}}\big) \, &\\
        + \, \mathcal{F} \, \circ \, \mathcal{E} \Big( \mathcal{G}_{\theta_{t+1}} \big(\mathcal{R} \, & \circ \, \mathcal{F}^{-1} (\vec{y}_{t}) \big) \Big),
    \end{split}
    \label{eq:end2endvarnet}
\end{equation}
where $\vec{y}_{t} \, = \, (\vec{y}_t^1, \, ...,\, \vec{y}_t^{n_c}) \,:= \,\mathcal{F} \, \circ \, \mathcal{E}(\vec{z}_{t})$. Note that for the conversion of \eqref{eq:varnet} into \eqref{eq:end2endvarnet} we used the facts $$\mathcal{F} \, \circ \, \mathcal{E} \, \circ \, \mathcal{R}  \, \circ \, \mathcal{F}^{-1}\,=\,\mathbb{1}_{\mathbb{C}^{n\times n_c}},\quad \mat{U}\,\circ\,\mat{U}\,=\,\mat{U}.$$ 
% The authors of E2EVarNet used as the image-refinement model $\mathcal{G}_{\theta_t}$ a U-Net \cite{ronneberger2015unet} in each layer.

% by adapting the recurrent ideas of Recurrent Inference Machines, 
Lastly, we replaced the image-refinement model $\mathcal{G}_{\theta_t}$ with a recurrent unit in every layer, which we further describe in the rest of this Section. 

\begin{figure}
    \centering
    \includegraphics[width=1.0\linewidth]{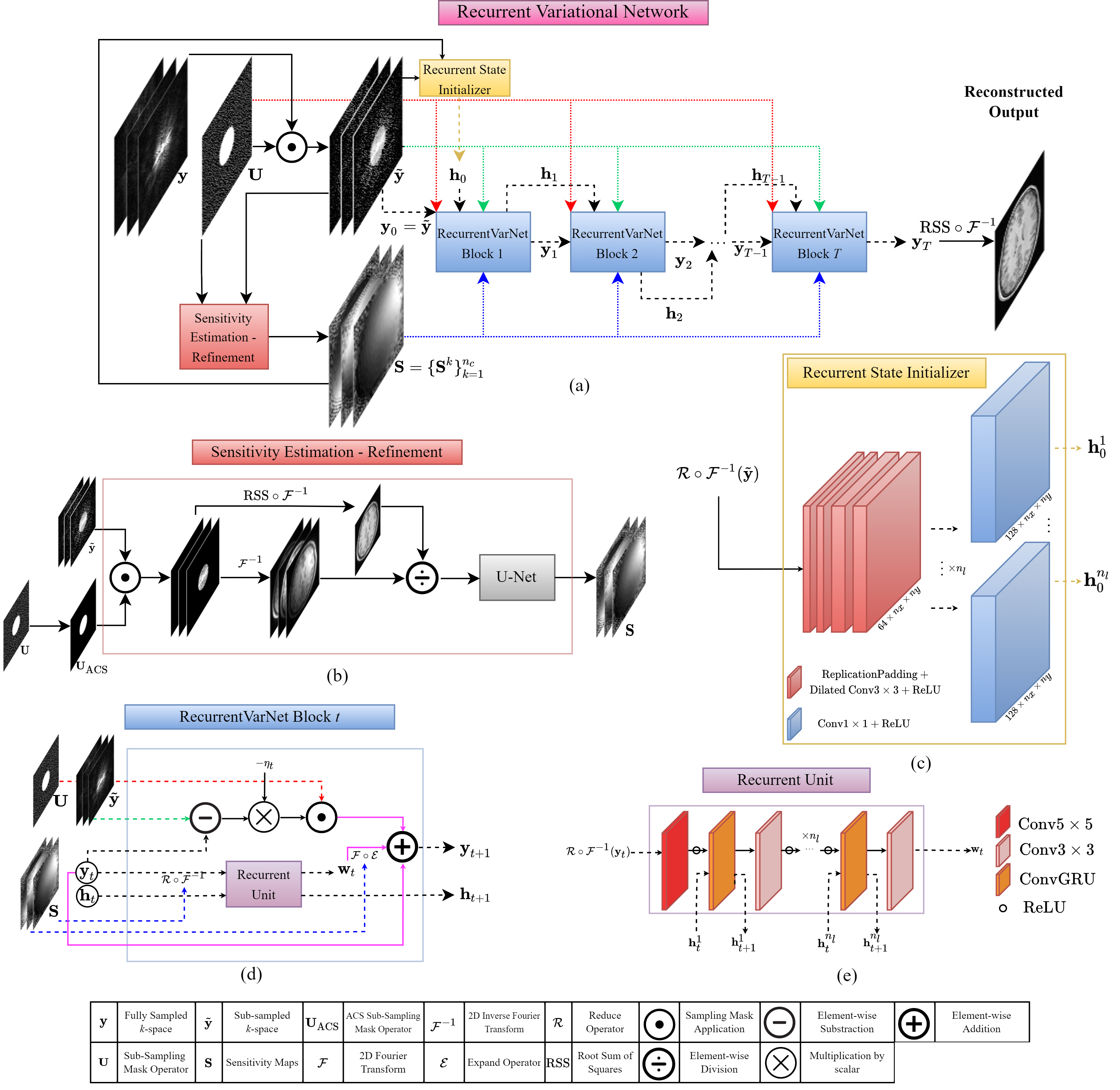}
    \caption{(a) Diagram of the end-to-end training pipeline of our proposed \textbf{Recurrent Variational Network}. During training the fully sampled multi-coil $k$-space data $\vec{y}$ are sub-sampled into $\tilde{\vec{y}}$ and fed to the model which iteratively refines $\tilde{\vec{y}}$ over $T$ time-steps. The final $k$-space prediction is projected onto the image domain and reconstructed into the output via the RSS operator. (b) The \textbf{Sensitivity Estimation - Refinement (SER)} module estimates and refines the coil sensitivity maps. (c) The \textbf{Recurrent State Initializer (RSI)} outputs the hidden state for the first RecurrentVarNet Block based on the SENSE reconstruction of the sub-sampled data. (d) The \textbf{RecurrentVarNet Block} is the main block of the RecurrentVarNet. Each block at iteration $t$ is fed intermediate quantities of the hidden state and the predicted $k$-space which it refines. (e) The \textbf{Recurrent Unit} of the RecurrentVarNet Block. It is consisted of alternating convolutions and ConvGRUs. 
    }
\label{fig:overview}
\end{figure}
\subsection{Recurrent Variational Networks}
\label{sec:subsec3.1}

The RecurrentVarNet takes as input a multi-coil sub-sampled $k$-space $\tilde{\vec{y}}$ and outputs a reconstructed prediction of the ground truth image. 
It consists of the main block, Recurrent Variational Network Block (RecurrentVarNet Block),  a Sensitivity Estimation and Refinement (SER) module,  and a Recurrent State Initializer (RSI) module. The RecurrentVarNet Blocks produce intermediate $k$-space refinements of $\tilde{\vec{y}}$ following an unrolled iterative scheme of $T$ time-steps.  The SER module estimates and refines (during training) the sensitivity map of each coil and the RSI module produces an initialization of the hidden or internal state of the first RecurrentVarNet Block. The above will be further defined in Sections \ref{sec:subsec3.2}, \ref{sec:subsec3.4} and \ref{sec:subsec3.3}, respectively. 

Note that our model uses as an initial guess of the $k$-space the sub-sampled $k$-space measurements $\vec{y}_0\,=\,\tilde{\vec{y}}$. In Fig. \ref{fig:overview} we provide a comprehensive illustration of the architecture of our RecurrentVarNet. 

\subsection{Recurrent Variational Network Block}
\label{sec:subsec3.2}
In this Section we formally introduce the main block of our proposed method: the Recurrent Variational Network Block. Following \eqref{eq:end2endvarnet} we replaced $\mathcal{G}_{\theta_{t}}$ with a Recurrent Unit $\mathcal{H}_{\theta_{t}}$. The iterative optimization scheme of the RecurrentVarNet performed by the RecurrentVarNet Block (\Fig{overview}(d)) has the following form 
\begin{equation}
    \begin{gathered}
        \vec{w}_t,\,\vec{h}_{t+1} \, = \, \mathcal{H}_{\theta_{t+1}} \Big(\mathcal{R} \, \circ \, \mathcal{F}^{-1} \big(\vec{y}_{t}\big),\, \vec{h}_{t} \Big)\\
        \
        vec{y}_{t+1} \, = \, \vec{y}_{t} \,- \, \alpha_{t+1} \,\mat{U}\big(\vec{y}_{t}\,-\,\tilde{\vec{y}}\big) \, + \, \mathcal{F} \, \circ \, \mathcal{E} \big( \vec{w}_{t} \big).       \\
    \end{gathered}
    \label{eq:rvarnet}
\end{equation}
Equation \ref{eq:rvarnet} comprises a data consistency term and a $k$-space refinement term. The latter projects the intermediate $k$-space prediction onto the image domain which in turn is fed to $\mathcal{H}_{\theta_{t}}$, whose output is projected back to the measurements domain, making RecurrentVarNet a hybrid domain-learning model. 

Contrary to other CNN-based architectures and similarly to recurrent neural networks, at each time-step $t$ of the unrolled optimization scheme our recurrent unit $\mathcal{H}_{\theta_t}$ takes additionally as input a hidden state $\vec{h}_{t-1}$ that stores the sequence information up to time-step $t-1$  and outputs the next $\vec{h}_t\,=\,(\vec{h}_t^1,\,...,\,\vec{h}_t^{n_l})$. The initial hidden state $\vec{h}_0$ is produced by the learned RSI unit (introduced in \Sec{subsec3.3}). 

As depicted in \Fig{overview}(e), the recurrent unit $\mathcal{H}_{\theta_t}$ of RecurrentVarNet Block consists of a two-dimensional $5\times 5$ convolution (Conv$5\times 5$), followed by a sequence of $n_l$  alternating layers of two-dimensional $3\times 3$ convolutions (Conv$3\times 3$) and ConvGRUs \cite{convrnn2017}. A ReLU activation function is applied after each convolution except the last. At each time-step $t$, each ConvGRU at a layer $i\,=\,1,\,...,\,n_l$ uses $h_{t-1}^{i}$ as an internal state produced from the corresponding layer $i$ of the previous time-step and outputs $h_{t}^{i}$ for the next time-step. In addition, for each block a trainable parameter posing as the step size $\alpha_{t}\,=\,\alpha_{\theta_t}$ at each iteration is learned which is initialized as 1.0. 

\subsection{Recurrent State Initializer}

Each RecurrentVarNet Block is fed with the previous internal state and outputs the next. In our framework we employed a Recurrent State Initializer (RSI) that learns to initialize the hidden state $\vec{h}_0$ inputted in the first RecurrentVarNet Block based on the sub-sampled input $\vec{y}_0\,=\,\tilde{\vec{y}}$. As an input to the RSI module we used a SENSE reconstruction
\begin{equation}
    \mathcal{R}\circ\mathcal{F}^{-1}(\vec{y}_0)\,=\,\sum_{k=1}^{n_c}{\mat{S}^k}^*\mathcal{F}^{-1}(\vec{y}_0^k).
    \label{eq:rsiinput}
\end{equation} 
As an alternative to \eqref{eq:rsiinput}, a zero-filled root-sum-of-squares reconstruction can be used
\begin{equation}
    \text{RSS}\circ\mathcal{F}^{-1}(\vec{y}_0)\,=\ \text{RSS}\big(\mathcal{F}^{-1}(\vec{y}_0^1),\,...,\,\mathcal{F}^{-1}(\vec{y}_0^{n_c})\big),
\end{equation}
but our experiments have shown superior performance when using the SENSE input.

Our Recurrent State Initializer was inspired by the works in \cite{yu2016multiscale} and it incorporates a sequence of four layers of a replication padding (ReplicationPadding) module \cite{liu2018partial} of size $p$ followed by a two-dimensional $3\times 3$ dilated convolution with $p$ dilations and $n_d$ filters.  Subsequently, the output  is fed into $n_l$ two-dimensional $1\times 1$ convolutions followed by a ReLU activation to produce the initial hidden state $\vec{h}_0\,=\,(\vec{h}_0^1,\,...,\,\vec{h}_0^{n_l})$.  Figure \ref{fig:overview}(c) provides an illustration of RSI's architecture. 
% For the values of $p$ and $n_d$ in each of the four layers we use (1, 1, 2, 4) and (32, 32, 64, 64), respectively.
\label{sec:subsec3.3}

\subsection{Sensitivity Estimation - Refinement}
\label{sec:subsec3.4}

For highly sub-sampled data using only the auto-calibration method as described by \eqref{eq:sensemaps} fails to produce accurate estimations of the coil sensitivity maps. For that reason, to estimate and refine the sensitivity map of each coil the RecurrentVarNet employs a Sensitivity Estimation - Refinement (SER) module which is jointly trained with the rest of the blocks of the model. Specifically, SER takes as input the sub-sampled $k$-space $\tilde{\vec{y}}$ and estimates an initial approximation of the sensitivity maps as in \eqref{eq:sensemaps}. Afterwards, this initial approximation is fed into a U-Net (denoted as $\mathcal{S}_{\theta_{s}}$) which refines the sensitivity maps even further:
\begin{equation}
    \mat{S}^k\,=\,\mathcal{S}_{\theta_{s}}(\tilde{\mat{S}}^k).
\end{equation}
An overview of the SER module is depicted by \Fig{overview}(b). Our U-Net implementation uses leaky-ReLU as activation instead of ReLU and performs instance normalization after each convolution.
%  For the U-Net $\mathcal{S}_{\theta_{s}}$ we use four layers  in the encoder and decoder with 8, 16, 32 and 64 number of filters for the convolutions, and zero dropout probability.

\subsection{Training Loss Function}
\label{sec:subsec3.5}

As aforementioned, our proposed model is fed with the sub-sampled $k$-space $\tilde{\vec{y}}$ as input and outputs at the last iteration a prediction $\vec{y}_T$ of the fully sampled $k$-space. For the reference or ground truth image we utilized the RSS reconstruction using the fully sampled measurements  $\vec{y}$ calculated as $\bar{\vec{x}}\,=\,\text{RSS}\circ\mathcal{F}^{-1}(\vec{y})$.

Our model is trained end-to-end. As a training loss function we employed a custom-made function $\tilde{\mathcal{L}}$ as the sum of the $\mathcal{L}_1$ loss and a loss derived from the Structural Similarity Index Measure (SSIM) metric \cite{1284395}, denoted as $\mathcal{L}_{\text{SSIM}}$. The loss was calculated as follows:
\begin{equation}
    \begin{gathered}
    \tilde{\mathcal{L}}( \bar{\vec{x}},\,\vec{x}_T) \, = \,w_1\mathcal{L}_1(\bar{\vec{x}},\,\vec{x}_T)\,+w_2\,\mathcal{L}_{\text{SSIM}}(\bar{\vec{x}},\,\vec{x}_T)\\
     = \,w_1 ||\bar{\vec{x}}\,-\,\vec{x}_T||_1\,+\,w_2(1-\text{SSIM}(\bar{\vec{x}},\,\vec{x}_T))
    \end{gathered}
\end{equation}
where 
\begin{equation}
    % \tilde{\vec{x}}\,=\,\text{RSS}\big(\mathcal{F}^{-1}(\tilde{\vec{y}}^1),\,...,\,\mathcal{F}^{-1}(\tilde{\vec{y}}^{n_c}) \big)
 \vec{x}_T\,=\,\text{RSS}\circ\mathcal{F}^{-1}(\vec{y}_T),
\end{equation}
and $0\le w_1,\,w_2 \le 1$ are multiplying factors.
\section{Experiments}
\label{sec:experiments}

\subsection{Implementation}
\label{sec:subsec4.1}

\subsubsection{RecurrentVarNet Implementation}
\label{sec:subsec4.1.1}
To avoid problems occurring from holomorphic differentiation of complex-valued data, we converted all complex-valued data to real-valued by concatenating the real and imaginary parts into the channels (= 2) dimensions. For instance, the masked $k$-space had the following form:
\begin{equation*}
    \Tilde{\Vec{y}}\,=\,\big( \text{Re}(\Tilde{\Vec{y}}),\,\text{Im}(\Tilde{\Vec{y}}) \big)\,\in\,\mathbb{R}^{2\times n \times n_c}.
\end{equation*}
Similarly, all operators and models operated in the real-valued domain. For instance,
\begin{equation*}
    \mathcal{A} : \mathbb{R}^{2\times n} \rightarrow \mathbb{R}^{2\times n\times n_c}, \quad \mathcal{A}^* : \mathbb{R}^{2\times n \times n_c} \rightarrow \mathbb{R}^{2\times n}.
\end{equation*}
With the above modification, our model works with multi-coil $k$-space data of arbitrary number of coils. Therefore, there is no need for prerequisite knowledge of the type of the MRI scanner.

Our model was implemented using PyTorch \cite{pytorch}. Our code and models including all state of the art baselines are available under the Apache 2.0 licence in our Deep Image Reconstruction Toolkit (DIRECT) \cite{DIRECTTOOLKIT}. For our experiments we used $T\,=\,8$ optimization steps (and RecurrentVarNet Blocks) and chose 128 channels for the size of the hidden state. Also, we set $n_l\,=\,4$ for the recurrent unit of RecurrentVarNet Block. For the RSI module as the values of $p$ and $n_d$ in each of the four layers we used (1, 1, 2, 4) dilations and (32, 32, 64, 64) filters, respectively, and used convolutions with 128 filters in the output layer. For $\mathcal{S}_{\theta_{s}}$ in the SER module we employed a U-Net with four scales with 8, 16, 32 and 64 number of filters for the convolutions, zero dropout probability, and 0.2 negative slope coefficient for the leaky-ReLU activation.

\subsubsection{Dataset}
\label{sec:subsubsec4.1.2}
\begin{figure}[ht!]
\begin{center}
\includegraphics[width=0.7\linewidth]{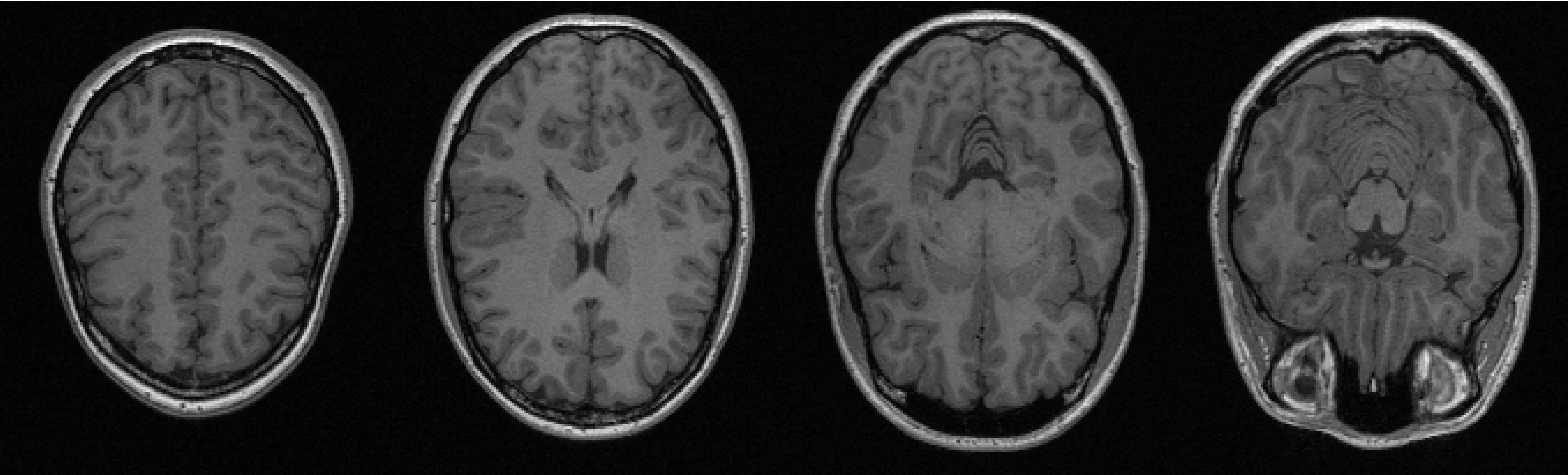}
\end{center}
\caption{Reconstructed  slices  with the RSS method of fully-sampled brain slices               from the Calgary-Campinas brain dataset.}
\label{fig:brain_samples}
\end{figure}

For our experiments we utilized the public Calgary-Campinas dataset \cite{calgary} which was released as part of the Multi-Coil MRI (MC-MRI) Reconstruction Challenge \cite{beauferris2020multichannel}. The dataset contains three-dimensional volumes of raw fully-sampled $k$-space measurements which are T1-weighted, gradient-recalled echo, 1 mm isotropic sagittal, acquired on 1.5T or 3T MRI scanners. 

We employed only the provided training (47 volumes, 7332 axial slices) and validation (20 volumes, 3120 axial slices) sets of the dataset (fully-sampled test set is not yet publicly available) which amounted to 67 volumes of $n_c = 12$ multi-coil data. We split the validation set in half to create a new validation set (10 volumes of 1560 axial slices) and a test set (10 volumes of 1560 axial slices). Reconstructed images have $218\times$ $170$/$180$ pixels. Some instances of reconstructed slices are illustrated in \Fig{brain_samples}.

\subsubsection{Sub-sampling}
\label{sec:subsubsec4.1.3}

We retrospectively sub-sampled the fully-sampled data setting as acceleration factors $R = 5$ and $R=10$ and using sub-sampling masks provided by the Calgary-Campinas MC-MRI Reconstruction Challenge.

\begin{figure}[ht!]
    \centering
    \includegraphics[width=0.15\linewidth]{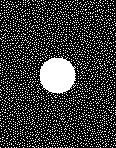}
    \caption{Sub-sampling mask}
\label{fig:mask}
\end{figure}

Figure \ref{fig:mask} depicts an example of a sub-sampling mask used. During training, data were arbitrarily 5-fold or 10-fold sub-sampled, and for evaluation and testing, data were both 5 and 10-fold sub-sampled.  

\subsubsection{Training Details}
\label{sec:subsubsec4.1.4}
Our model was optimized using the Adam optimizer with parameters $\beta_1\, = \,0.9$, $\beta_2 \,=\, 0.999$ and $\epsilon\,=\,1\text{e}-8$ with a batch size of 4 two-dimensional slices  for 63000 iterations which amounted to approximately 270 epochs. A warm-up schedule \cite{goyal2018accurate} was used to linearly increase the learning rate to 0.0005 over 1000 warm-up iterations. The learning rate was decayed by a factor of 0.2 every 20000 training iterations. For the training we employed eight NVIDIA RTX A6000 GPUs. 

For the training loss function we set both multiplying factors to $w_1,\,w_2\,=\,1.0$.  The total number of trainable parameters for our model with the choice of hyper-parameters as described in \Sec{subsec4.1.1} amounted to approximately 7400k parameters.

\begin{figure}[ht!]
    \centering
    \includegraphics[width=1.0\linewidth]{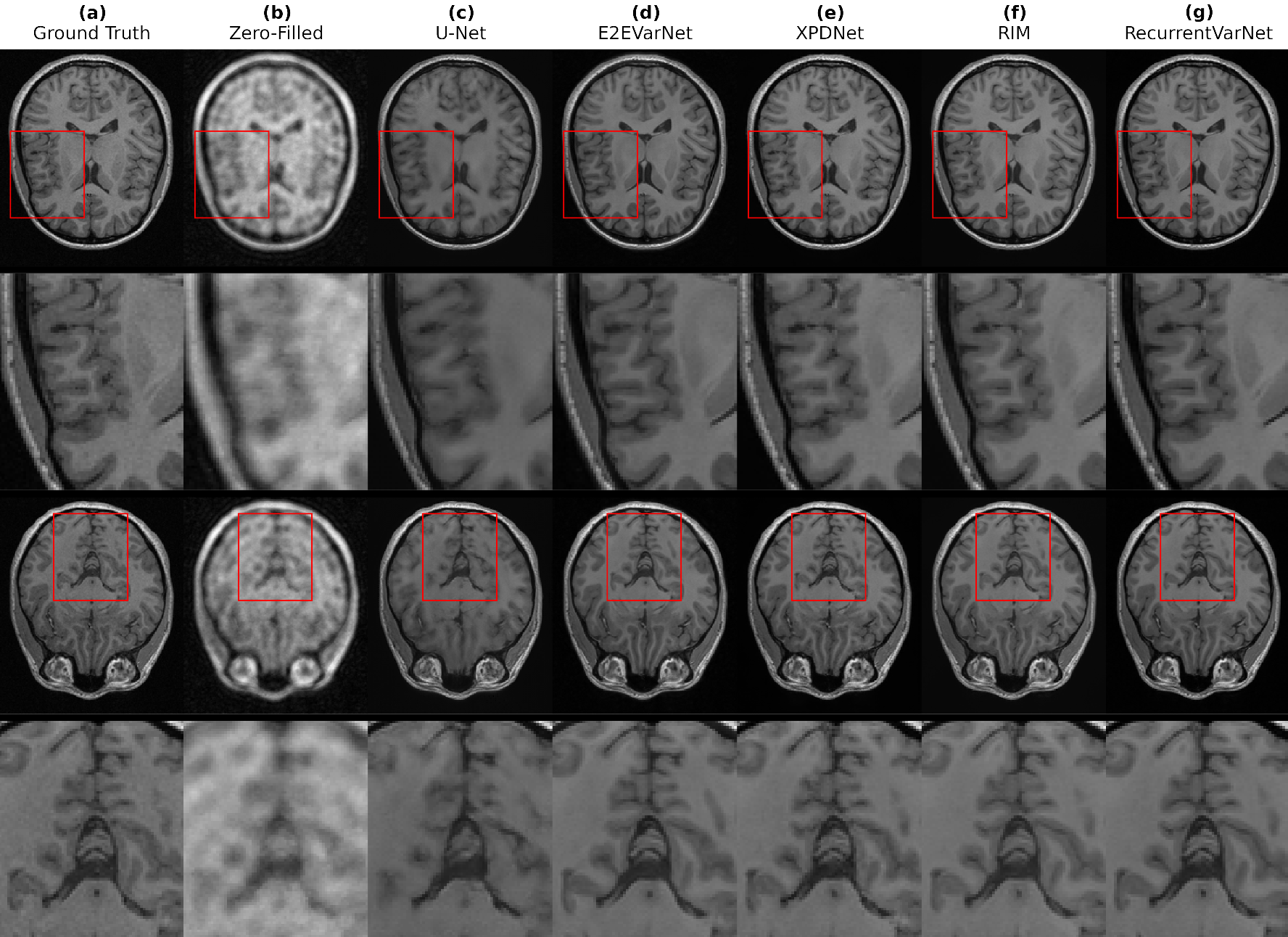}
    \caption{ Reconstructed slices from the test set along with zoomed-in regions of interest using an acceleration factor of $R\,=\,5$. (a) Ground Truth, (b) Zero-filled reconstruction, (c) U-Net, (d) End-to-End Variational Network, (e) XPDNet, (f) Recurrent Inference Machine, (g) Ours: \textbf{Recurrent Variational Network}.
    }
\label{fig:recons}
\end{figure}

\subsubsection{Evaluation Metrics}
\label{sec:subsubsec4.1.5}
To evaluate the quality of the reconstructions we employed three commonly used in image reconstruction assessment metrics: the SSIM metric, the peak signal-to-noise ratio (pSNR) \cite{5596999}, and the normalized mean squared error (NMSE) \cite{nmse}. The metrics were computed using the reference RSS reconstructions of the fully-sampled data and the predicted RSS reconstructions.

\subsection{Experimental Results}
\label{sec:subsec4.2}

%%%%%%%%%%%%%----SOTA TABLE----%%%%%%%%%%%%%%%%%%%%%%%%%%% 
\begin{table}[ht!]
\centering

\begin{tabular}{ccccc}
\hline
\hline
\multirow{2}{*}{Acceleration Factor} & \multirow{2}{*}{Model}   & \multicolumn{3}{c}{Metrics}                           \\ \cline{3-5} 
                                     &                          & SSIM $\uparrow$ & pSNR $\uparrow$ & NMSE $\downarrow$ \\ \hline \hline
\multirow{5}{*}{$R\,=\,5$}           & Zero-Filled              & 0.7371          & 25.27           & 0.0401            \\
                                     & U-Net                    & 0.8682          & 29.25           & 0.0210            \\
                                     & E2EVarNet                & 0.9073          & 32.62           & 0.0086            \\
                                     & XPDNet                   & 0.9325          & 34.80           & 0.0051            \\
                                     & RIM                      & 0.9381          & 35.61           & 0.0042            \\
                                     & \textbf{RecurrentVarNet} & \textbf{0.9418} & \textbf{36.02}  & \textbf{0.0038}   \\ \hline
\multirow{5}{*}{$R\,=\,10$}          & Zero-Filled              & 0.6905          & 24.35           & 0.0527            \\
                                     & U-Net                    & 0.8177          & 27.23           & 0.0308            \\
                                     & E2EVarNet                & 0.8610          & 29.85           & 0.0158            \\
                                     & XPDNet                   & 0.8949          & 31.76           & 0.0105            \\
                                     & RIM                      & 0.9047          & 32.52           & 0.0089            \\
                                     & \textbf{RecurrentVarNet} & \textbf{0.9073} & \textbf{32.71}  & \textbf{0.0085}   \\ \hline \hline
\end{tabular}
\caption{
Quantitative evaluation using three evaluation metrics for two acceleration factors $R\,=\,5,\,10$.
} % \caption
\label{tab:sota}
\end{table}
%%%%%%%%%%%%%%%%%%%%%%%%%%%%%%%%%%%%%%%%%%%%%%%%%%%%%%%%%%%
\subsubsection{Comparisons}
\label{sec:subsubsec4.2.1}

To evaluate the performance of our proposed model we made comparisons against reconstructions output from the following models:
\begin{itemize}[leftmargin=*]
\setlength\itemsep{-.3em}
    \item A U-Net with four scales consisted of 64, 128, 256, and 512 convolution filters respectively.
    \item An End-to-End Variational Network with the same choice of hyper-parameters as in the published work \cite{sriram2020endtoend}.
    \item An XPDNet \cite{ramzi2021xpdnet} with 10 iterations, 5 primal and 5 dual layers and U-Nets for the primal and dual models.
    \item A Recurrent Inference Machine with 16 time-steps and 128 features \cite{LONNING201964}.
\end{itemize}
All models were trained on the training set by arbitrarily setting $R$ to 5 or 10, and were optimized in similar fashion to \Sec{subsubsec4.1.4}. Additionally, all models were trained jointly with a U-Net that estimated and refined the sensitivity maps with the same choice of hyper-parameters as our SER module.  We also compared RecurrentVarNet's reconstructions against  Zero-Filled reconstructions (we used $\text{RSS}\circ\mathcal{F}^{-1}(\Tilde{\vec{y}})$). In the following paragraphs we provide quantitative and qualitative results and comparisons.

\paragraph{Quantitative Results}
\label{para:quant}

In \Tab{sota} are presented the quantitative results of our experiments for both acceleration factors and are reported the average evaluation metrics obtained on the test set. To obtain these results we used the best-performing checkpointed model according to the validation results on the SSIM metric. Compared against the other methods, RecurrentVarNet marked the highest average SSIM and pSNR metrics, as well as the lowest average NMSE metric, highlighting its superior performance.

\paragraph{Qualitative Results}
\label{para:qaul}

To asses the quality of the RecurrentVarNet's reconstructions we visualized in \Fig{recons} two reconstructed samples with the RSS method from the test set which were 5-fold sub-sampled. For comparisons, we also visualized the ground truth images, the images reconstructed from the sub-sampled data (Zero-Filled), and the trained models U-Net, E2EVarNet, XPDNet, and RIM. In contrast with the rest of the reconstructions, our method produced images that matched the ground truth images better as shown by the zoomed-in regions of interest in \Fig{recons}. Visualizations for $R\,=\,10$ can be found in the Supplementary Material.

\subsubsection{Ablation Study}
\begin{table}[ht!]
\centering
\begin{tabular}{ccccc}
\hline \hline
\multirow{2}{*}{Acceleration Factor} & \multirow{2}{*}{RecurrentVarNet}                         & \multicolumn{3}{c}{Metrics}                           \\ \cline{3-5} 
                                     &                                                & SSIM $\uparrow$ & pSNR $\uparrow$ & NMSE $\downarrow$ \\ \hline \hline
\multirow{5}{*}{$R\,=\,5$}           & No SER module                                  & 0.9374          & 35.52           & 0.0043             \\
                                     & Shared weights                                 & 0.9386          & 35.66         & 0.0041             \\
                                     & No RSI module                                  & 0.9411          & 35.93           & 0.0039               \\
                                     & \multicolumn{1}{l}{$T\,=\,11,\quad n_l\,=\,3$} & 0.9417          & 35.99           & 0.0037            \\
                                     & \textbf{Original}                       & \textbf{0.9418}  & \textbf{36.02} & \textbf{0.0038}   \\ \hline
\multirow{5}{*}{$R\,=\,10$}          & No SER module                                  & 0.8960          & 32.04           & 0.0099            \\
                                     & Shared weights                                 & 0.9017           &  32.37          & 0.0092                  \\
                                     & No RSI module                                  & 0.9071           &  32.69         &  0.0085                 \\
                                     & \multicolumn{1}{l}{$T\,=\,11,\quad n_l\,=\,3$} & \textbf{0.9093}  & \textbf{32.83} & \textbf{0.0082}   \\
                                     & Original                                       &    0.9073        & 32.71          & 0.0084            \\ \hline \hline
\end{tabular}
\caption{
Ablation Study of RecurrentVarNet: Quantitative results of $5$-fold and $10$-fold accelerated reconstructions using four distinct implementations of the RecurrentVarNet.
} % \caption
\label{tab:ablations}
\end{table}
To investigate the performance of our RecurrentVarNet even further, we compared the original implementation with different choices of hyper-parameters. More specifically, we compared it against four distinct cases:
\begin{itemize}[leftmargin=*]
\setlength\itemsep{-.3em}
    \item A RecurrentVarNet without a SER module. In this case, the sensitivity maps were estimated as in \eqref{eq:sensemaps} with no refinement performed.
    \item A RecurrentVarNet without a RSI module. The hidden state was initialized as a vector of zeros.
    \item A RecurrentVarNet with the same number of RecurrentVarNet Blocks, but with shared weights for the Recurrent Unit that is, $\mathcal{H}_{\theta_t}\,=\,\mathcal{H}_{\theta}$ for all $t\,=\,1,\,...,\,T$.
    \item A RecurrentVarNet with higher number of RecurrentVarNet Blocks ($T\,=\,11$) and less layers in the Recurrent Unit ($n_l\,=\,3$). 
\end{itemize}

For the ablation study due to space limitation we chose to present here only the quantitative comparisons of the original implementation with the four modifications, and we include the qualitative comparisons in the Supplementary Material. Similarly to \Sec{subsubsec4.2.1}, we utilized the best model checkpoints to acquire the evaluation metrics. In \Tab{ablations} we report the average SSIM, pSNR and NMSE metrics acquired from the reconstructions of the 5-fold and 10-fold accelerated test set. 

In general, our original model exhibited superior performance for both acceleration factors, therefore, our original implementation is well-justified. It should be noted that the RecurrentVarNet with the choice of $T\,=\,11$ and $n_l\,=\,3$ performed slightly better in the case of $R\,=\,10$, which can be attributed to the fact that more optimization steps were used. However, our original architecture was the model of choice since the model with $T\,=\,11$ and $n_l\,=\,3$ required more memory during training due to its need to accumulate more gradients in memory to perform the back-propagation when computing the loss function. 
\section{Conclusions}
\label{sec:sec5}

In this work, we have proposed a novel deep learning image inverse problem solver architecture, the Recurrent Variational Network, which we particularized to the task of reconstructing MR images from accelerated multi-coil $k$-space measurements. The quantitative and qualitative results of our experiments have demonstrated that the RecurrentVarNet outperformed state of the art approaches, which can be attributed to its hybrid nature, since its main block performs iterative  optimization in the $k$-space domain using a recurrent unit operating in the image domain.

%Bibliography
\bibliographystyle{unsrt}  
\bibliography{references}  

\newpage
% % --- PDF will be split by an editor (e.g. macOS preview), so need to restart from page 1
% \setcounter{page}{1}

% \onecolumn
% \twocolumn[
\centering
\begin{Large}
\textbf{Recurrent Variational Network: A Deep Learning Inverse Problem Solver applied to the task of Accelerated MRI Reconstruction} \\
\vspace{1.0em}Supplementary Material \\
\end{Large}

\appendix

\begin{flushleft}

\label{appendix}
\section{Additional Figures}
\subsection{Comparisons - Qualitative Results}
\label{appendix:comparisons}
\end{flushleft}

\begin{figure*}[ht!]
    \centering
    \includegraphics[width=1.0\linewidth]{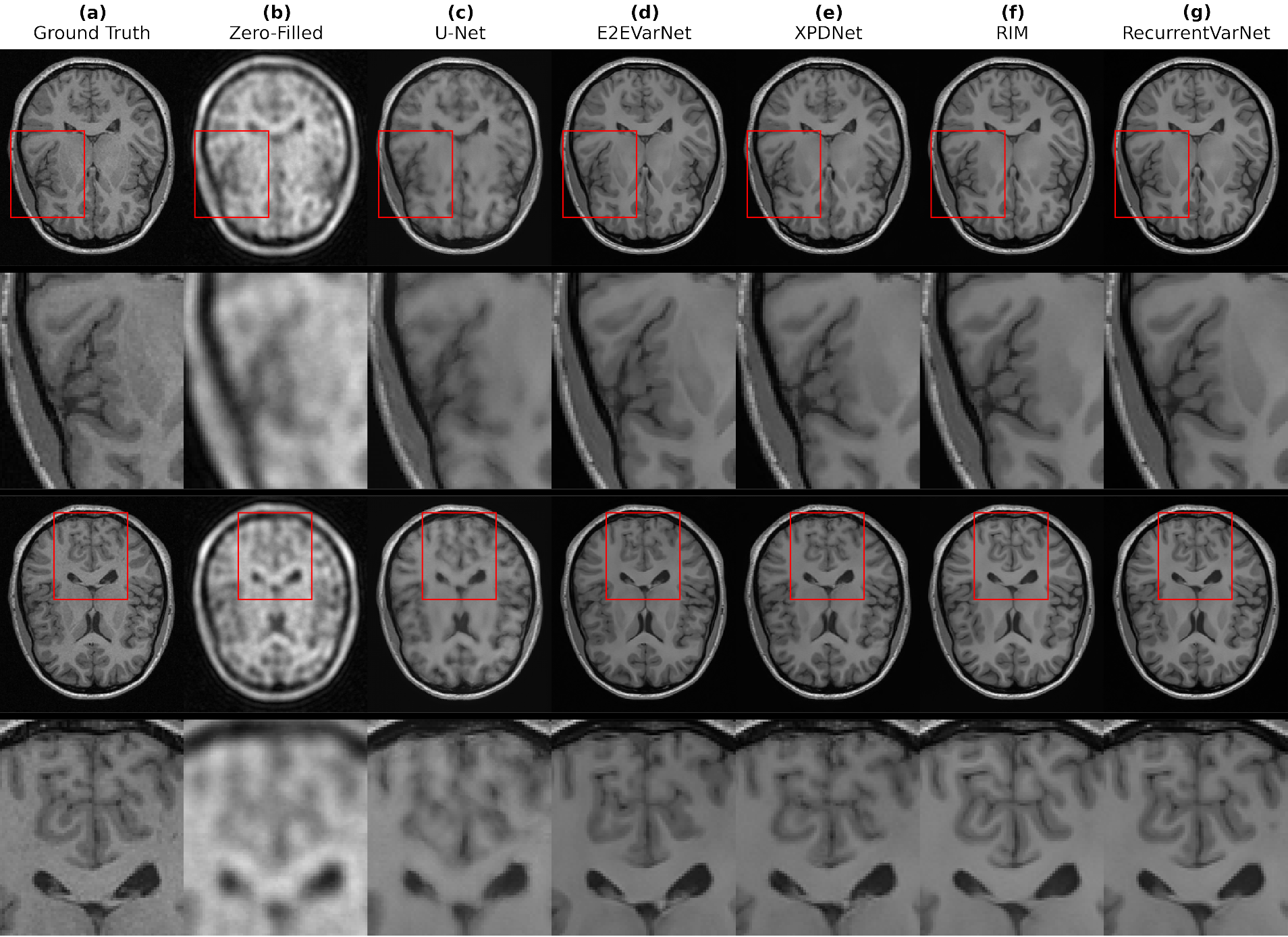}
    \caption{ Reconstructed slices from the test set along with zoomed-in regions of interest using an acceleration factor of $R\,=\,10$. (a) Ground Truth, (b) Zero-filled reconstruction, (c) U-Net, (d) End-to-End Variational Network, (e) XPDNet, (f) Recurrent Inference Machine, (g) Ours: \textbf{Recurrent Variational Network}.
    }
\label{fig:recons_10x}
\end{figure*}

\newpage
\begin{flushleft}
\subsection{Ablation Study - Qualitative Results}
\label{appendix:ablation}
\end{flushleft}

\begin{figure}[ht!]
    \centering
    \includegraphics[width=1.0\linewidth]{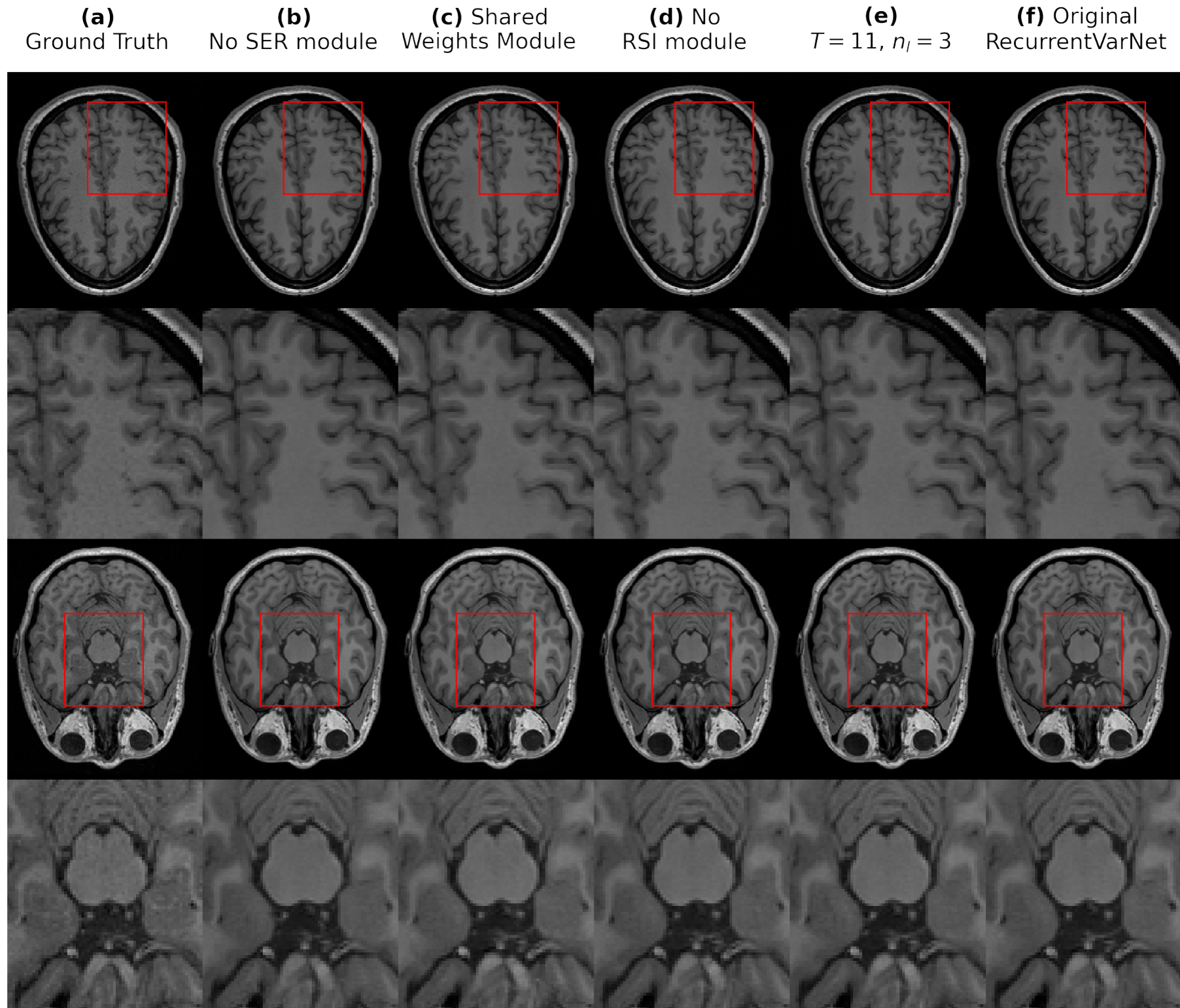}
    \caption{ Reconstructed slices from the test set along with zoomed-in regions of interest using an acceleration factor of $R\,=\,5$. (a) Ground Truth, (b) ReccurentVarNet without a Sensitivity Estimation - Refinement module, (c) ReccurentVarNet with shared weights for the Recurrent Units, (d) ReccurentVarNet without a Recurrent State Initializer, (e) ReccurentVarNet with $T\,=\,11$ and $n_l\,=\,3$, (f) Original Recurrent Variational Network.
    }
\label{fig:recons_ablation}
\end{figure}
\newpage
\begin{figure}[ht!]
    \centering
    \includegraphics[width=0.9\linewidth]{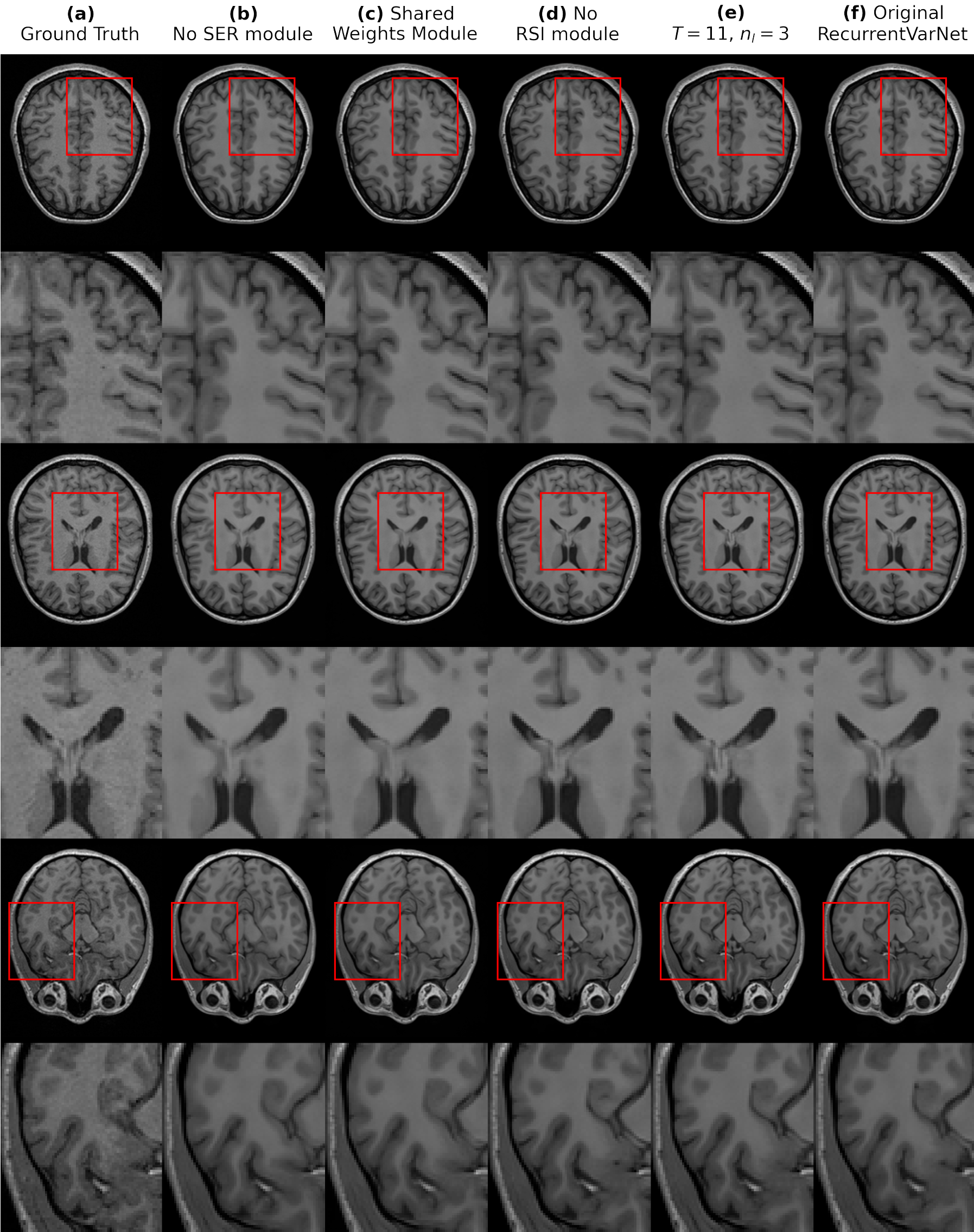}
    \caption{ Reconstructed slices from the test set along with zoomed-in regions of interest using an acceleration factor of $R\,=\,10$. (a) Ground Truth, (b) ReccurentVarNet without a Sensitivity Estimation - Refinement module, (c) ReccurentVarNet with shared weights for the Recurrent Units, (d) ReccurentVarNet without a Recurrent State Initializer, (e) ReccurentVarNet with $T\,=\,11$ and $n_l\,=\,3$, (f) Original Recurrent Variational Network.
    }
\label{fig:recons_ablation_10x}
\end{figure}
\newpage

\begin{flushleft}
\section{Additional Experiments}
\end{flushleft}
\begin{justify}
In this Section we provide additional experiments for our proposed method using the fastMRI AXT1 multi-coil brain dataset \cite{zbontar2019fastmri}.
\end{justify}

\begin{flushleft}
\subsection{Dataset description}
\label{sec:fatsmri_axt1}
\end{flushleft}

\begin{justify}
The fastMRI AXT1 multi-coil brain dataset is consisted of 248 training volumes (3844 slices) and 82 validation (1424 slices) validation volumes which we split in half to create new validation and test sets (41 volumes/712 slices each). 
\end{justify}

\begin{flushleft}
\subsection{Sub-sampling}
\label{sec:fatsmri_axt1_subsampling}
\end{flushleft}

\begin{justify}
The data were retrospectively sub-sampled for $R=4$ and $R=8$ using random Cartesian masks provided by the fastMRI challenge. The sub-sampling masks were produced by first sampling a fraction of the low frequencies or the ACS region (8\% when $R = 4$ and 4\% when $R = 8$) and then randomly sampling up to the level of acceleration $R$. An example of each mask is depicted in \Fig{masks}. 
\begin{figure}[ht!]
     \centering
     \begin{subfigure}[b]{0.4\textwidth}
         \centering
         \includegraphics[width=0.6\textwidth]{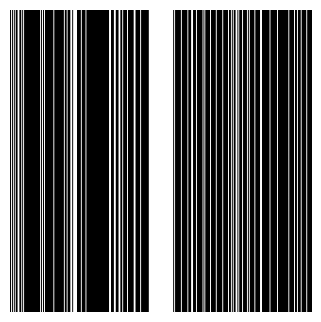}
         \caption{$R=4$ with 8\% of ACS region.}
         \label{fig:mask4x}
     \end{subfigure}
    %  \hfill
     \begin{subfigure}[b]{0.4\textwidth}
         \centering
         \includegraphics[width=0.6\textwidth]{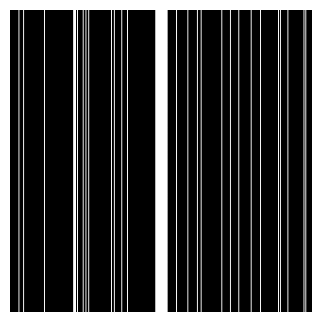}
         \caption{$R=8$ with 4\% of ACS region.}
         \label{fig:mask8x}
     \end{subfigure}
        \caption{Random Cartesian sub-sampling masks.}
        \label{fig:masks}
\end{figure}
\end{justify}

\begin{flushleft}
\subsection{Comparison \& Ablation Study}
\label{sec:comparison}
\end{flushleft}

\begin{justify}
We employed a RecurrentVarNet with $T=8$ time-steps, 92 filters for the hidden state $\mathbf{h}_t$, and $n_l=3$. For comparisons we used:
\begin{itemize}
    \item A U-Net with 32, 64, 128, 256 convolution filters in each scale.
    \item An E2EVarNet with 6 layers and 16, 32, 64, 128 filters for the scales of the U-Net regularizer.
    \item An XPDNet with 6 iterations and 5 primal and 5 dual layers with U-Nets  for the primal and dual models.
    \item A RIM with 8 time-steps and 64 features.
\end{itemize}

For the ablation study we employed a RecurrentVarNet with $T=6$ time-steps, 128 filters for $\mathbf{h}_t$, and $n_l=2$. 

All models were trained on four NVIDIA RTX A6000 GPUs. The RecurrentVarNet with $T=8$ and the XPDNet were trained for 120k iterations, while the rest of the aforementioned models were trained for 180k iterations.

Note that for our experiments on the fastMRI dataset we used a lower number of parameters than for those on the Calgary Campinas dataset due to the fact that the fastMRI slices have approximately $6\times$ more pixels and, hence, require approximately $6\times$ more GPU memory.

\paragraph{Quantitative results}
\label{sec:subsec_fastmri_axt1_quant}
\begin{table}[ht!]
\centering
% \resizebox{\linewidth}{!}{ %< auto-adjusts font size to fill line

\begin{tabular}{ccccccc}
\hline
\hline
\multirow{3}{*}{Model} & \multicolumn{6}{c}{Metrics}                                                                                                        \\ \cline{2-7} 
                       & \multicolumn{3}{c}{$R=4$}                                                  & \multicolumn{3}{c}{$R=8$}                             \\ \cline{2-7} \noalign{\vskip\doublerulesep
         \vskip-\arrayrulewidth} \cline{2-7}
                       & SSIM $\uparrow$ & pSNR $\uparrow$ & \multicolumn{1}{c|}{NMSE $\downarrow$} & SSIM $\uparrow$ & pSNR $\uparrow$ & NMSE $\downarrow$ \\ \hline \hline
    Zero-Filled            & 0.813                   & 30.77                    & \multicolumn{1}{c|}{0.039}             & 0.708           & 26.21           & 0.096             \\
    U-Net                  & 0.943                   & 37.09                    & \multicolumn{1}{c|}{0.012}             & 0.922           & 34.41           & 0.021             \\
    E2EVarNet              & 0.959                   & 40.83                    & \multicolumn{1}{c|}{0.005}             & 0.940           & 37.35           & 0.011             \\
    XPDNet                 & 0.958                   & 40.93                    & \multicolumn{1}{c|}{0.005}             & 0.941           & \textbf{37.83}           & 0.010             \\
    RIM                    & 0.957                   & 40.84                    & \multicolumn{1}{c|}{0.005}             & 0.938           & 37.36           & 0.011             \\
RecurrentVarNet        & \textbf{0.960}              & \textbf{41.28}           & \multicolumn{1}{c|}{\textbf{0.005}}    & \textbf{0.941}           & 37.69           & \textbf{0.010}             \\
RecurrentVarNet (ablation) & 0.959           & 40.95           & \multicolumn{1}{c|}{0.005}             & 0.938           & 37.31           & 0.011             \\ \hline \hline
\end{tabular}

\caption{
Quantitative evaluation using three evaluation metrics for two acceleration factors $R\,=\,4,\,8$ on the fastMRI AXT1 test set.
} % \caption
\label{tab:quant}
\end{table}
\\
In \Tab{quant} are reported the average evaluation metrics obtained on the test set  for both acceleration factors.

\paragraph{Qualitative results}
\\
In \Fig{recons_4x_fastmri} are visualized reconstructed slices with the RSS method from one sample from the test set for $R=4$.
\label{sec:subsec_fastmri_axt1_qual}
\begin{figure*}[ht!]
    \centering
    \includegraphics[width=1.0\linewidth]{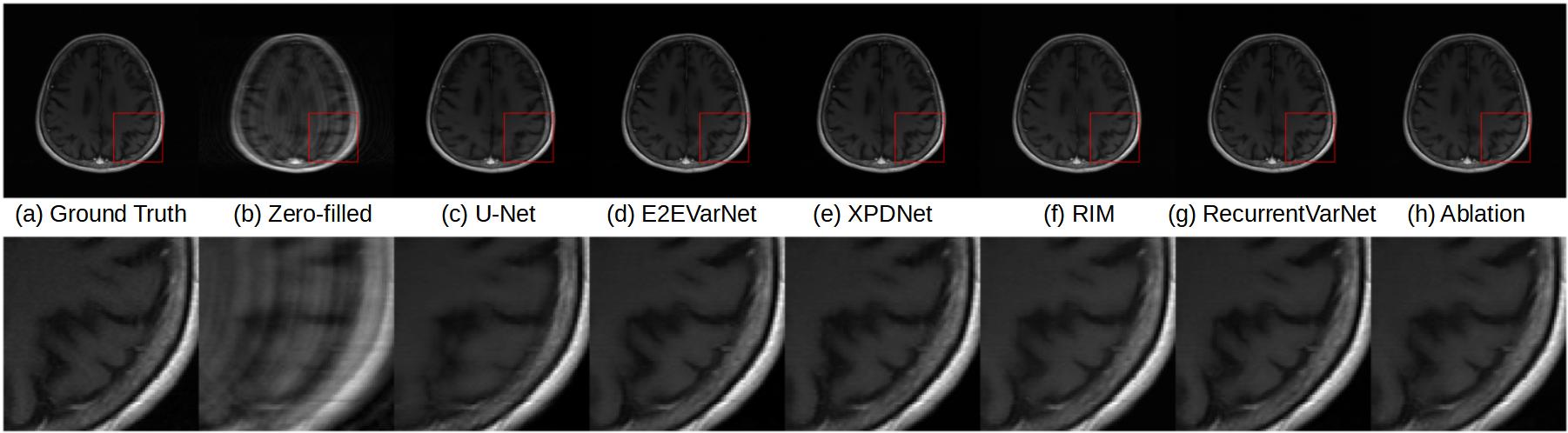}
    \caption{ Reconstructed slice from the fastMRI AXT1 test set along with zoomed-in regions of interest using an acceleration factor of $R\,=\,4$. (a) Ground Truth, (b) Zero-filled reconstruction, (c) U-Net, (d) End-to-End Variational Network, (e) XPDNet, (f) Recurrent Inference Machine, (g) Ours: Recurrent Variational Network, (h) Recurrent Variational Network (ablation).
    }
\label{fig:recons_4x_fastmri}
\end{figure*}

\end{justify}
%%%%%%%%% REFERENCES

\end{document}